\documentclass[
superscriptaddress,
amsmath,amssymb,
aps,
12pt]{revtex4}
\usepackage{float}
\usepackage{natbib}
\usepackage{mathtools}
\usepackage{xcolor}
\usepackage{eucal}
\usepackage[labelformat=simple]{subcaption}
\usepackage[dvipsnames]{xcolor}
\usepackage[utf8]{inputenc}
\usepackage{amsmath}
\usepackage[T1]{fontenc}
\usepackage{amsmath,amsfonts,amssymb,amsthm,epsfig,epstopdf,url,array}
\usepackage{soul}
\usepackage{braket}
\usepackage{dcolumn}% Align table columns on decimal pointnatbib
\usepackage{bm}% bold math
\usepackage{graphicx}
\usepackage{comment}
\usepackage{bm}
\usepackage[caption=false]{subfig}
\usepackage{lipsum}
\usepackage{braket}
\usepackage{soul}
\begin{document}
\title{Quantum backflow in biased tight-binding systems}
% Force line breaks with \\

\author{Francisco Ricardo Torres Arvizu\footnote{Email: ftorres@icf.unam.mx}}
\affiliation{Instituto de Ciencias Físicas, UNAM, Av. Universidad s/n, ZIP 62210 Cuernavaca Morelos, México}

\author{Adrian Ortega}
\affiliation{Centro Universitario de Guadalajara, Universidad de Guadalajara, Calle Guanajuato 1045, ZIP 44160, Guadalajara, Jalisco, México}

\author{Hernán Larralde}
\affiliation{Instituto de Ciencias Físicas, UNAM, Av. Universidad s/n, ZIP 62210 Cuernavaca Morelos, México}

\begin{abstract}
    % We present a study of the spectral and transport properties of tight binding systems with complex couplings, considering various boundary conditions and lattice sizes. The complex couplings give rise to behavior reminiscent to that of a classically biased diffusive systems. Interestingly, the phenomenon of quantum backflow is present. This is an intrinsically non-classical effect where the density flux associated with a particle described by the superposition of wave functions with positive momentum acquires negative values. We calculate the superposition that gives rise to the strongest backflow in the various systems.
    We study the phenomenon of quantum backflow in tight-binding systems with complex couplings, considering different boundary conditions and lattice sizes. Backflow is an intrinsically non-classical effect where the density flux associated with a particle described by the superposition of wave functions with, say, positive-momentum, acquires negative values. We calculate the superposition of positive-momentum states that gives rise to the strongest backflow in the  system. We also evaluate the bounds on the total amount of probability that flows in the opposite direction of the particle's momentum. 
\end{abstract}

 \maketitle

\section{Introduction}
The phenomenon of backflow is a non-classical and non-linear effect that occurs when the probability current (or density flux) of a particle takes values opposite to its momentum. This phenomenon was first identified in 1969 by Allcock in the context of arrival times in quantum mechanics \cite{ALLCOCK1969253}.  % when he observed that the arrival rate was not monotonically increasing with time. Later, in 1994
Bracken and Melloy \cite{Bracken_1994} systematically formulated the problem in terms of the probability current of a normalized wave packet with positive-momentum. The most significant outcome of their approach is that the total probability that flows in the direction opposite to the momentum is bounded by a universal constant $c_{\text{BM}}$. The exact value of the constant $c_{\text{BM}}$ is unknown, with the best numerical estimates obtained in~\cite{Penz_2006, eveson}, suggesting that in the infinite continuous case considered in those works, $c_{\text{BM}}\approx 0.0384517$. 

The phenomenon of backflow has been investigated in various extensions of the original setup \cite{Dissipative, Goussev1, Melloy1, Goussev2}, including relativistic generalizations \cite{Relativistic}, scattering conditions \cite{Scatering}, and many-particle formulations \cite{many_particle}. Backflow has also been reported in other systems, such as quasi-stable quantum systems \cite{Dijk}, Hermite wave packets \cite{Strange_2024}, and in connection with retropropagation and negative kinetic energy \cite{Berry_2010}.
Continuing in this vein, it is natural to extend the study of this phenomenon to discrete versions of such systems, and to explore how they differ from their continuous counterparts. In this work we focus on the nature of backflow in tight-binding chains with constant (complex) next nearest-neighbor couplings and onsite energies, which can be regarded as discrete analogues of a free-particle–type Hamiltonian ~\cite{Timothy}. 
The tight-binding model has traditionally been used to determine electronic and transport properties in solids with periodic potentials, such as insulators and semiconductors \cite{ashcroft2011solid}. However, its versatility and simplicity have allowed it to be used to characterize a wide variety of  physical systems, ranging from waveguides \cite{waveguide1, waveguide2, waveguide3, waveguide4} to quantum random walks \cite{qrw, qrw1}. We consider both periodic and infinite chains, which correspond to discrete versions of the general backflow setups of particles confined to a continuous infinite straight line \cite{Bracken_1994} and ring respectively \cite{ring1, Ring2}. Interest in periodic systems arises from several characteristics, among them that the momentum eigenfunctions are normalizable and their spectrum is discrete, making the calculation of the flux bounds easier to a certain extent \cite{ring1, Strange_2012}.  Other  generalizations using periodic boundary conditions have been employed to study backflow in a variety of situations, such as relativistic spin-$1/2$ particles \cite{dirac_1_2}, massless charged Dirac fermions \cite{DIBARI2023128831}, and systems of two identical particles \cite{TwoParticlesBarbier_2025}. Extensions to two-dimensional systems  \cite{Paccoia}, where it turns out that the backflow is not bounded \cite{twodimensions}, have also been considered. \\

Recently, complex couplings in one-dimensional tight-binding chains have attracted interest, among other features, for breaking time-reversal symmetry while the Hamiltonian remains Hermitian \cite{ComplexTB}. In the context of this work, from a transport perspective, their relevance lies in the fact that the imaginary part of the hopping parameter can be interpreted as a bias, analogous to that found in classical diffusive systems \cite{Zimbor, Lu, Liu}. As will be explained in more detail in the paper, this results in an additional term in the expression for the probability density flux. Considering this, we address two questions that naturally arise regarding quantum backflow: %(see, for example, \cite{Ring2}).These are:
How large can the backflowing probability density flux be? and second, along the lines of \cite{Bracken_1994}: Is the total probability flowing backwards bounded? Surprisingly, the first question has rarely been explored. Recently, Goussev \cite{Ring2} was among the first to delve into this subject,  obtaining the optimal bounds of the probability density flux for a finite superposition of low-energy eigenstates of the Hamiltonian of a particle confined to a ring.  In a similar vein, we construct wave packets composed exclusively of positive-momentum states that exhibit the most pronounced instantaneous negative flux for both infinite and periodic chains.  In this respect, it is worth noting that tight-binding systems, unlike their continuous counterparts, exhibit a bounded energy spectrum as a consequence of lattice discretization,  which implies that it is not necessary to restrict the number of admissible non-negative momentum eigenstates to avoid infinite energies \cite{Ring2}.  Moreover, the inclusion of a bias results in an enhancement of the flux amplitudes as its strength is increased, while simultaneously allowing controlled adjustments of the spectral window of suitable eigenstates. On the other hand,  concerning the second question, we find that, regardless of the system topology, the total integrated backflow in the simple tight-binding systems we consider, exceeds that of their continuous counterparts in certain parameter regions.  In this respect, both approaches highlight complementary but related aspects of quantum backflow: the first leads to a more pronounced but shorter-lived negative flux whereas the second yields an oscillatory flux characterized by negative-valued oscillations, albeit with a more moderate amplitude.\\

%In what follows we address two questions that naturally arise in regards to this phenomenon, the first of which, somewhat surprisingly, has been rarely explored (see, for example, \cite{Ring2}). These are: How large can the backflowing probability density flux be? and second, along the lines of \cite{Bracken_1994}: Is the total probability flowing backwards bounded?.
 
The outline of the paper is as follows: In Section II, we introduce the systems we will consider and present the solution of the corresponding Schrödinger equation. Section III is devoted to the continuity equations and to the physical role of the imaginary part of the coupling constant in these systems. In Section IV, we determine the momentum operator associated with the system, with the aim of identifying the conditions under which its eigenvalues are positive, thereby enabling the construction of superpositions of states with strictly positive-momentum. In Section V, we calculate the corresponding fluxes and analyze the optimal bounds of the probability current, considering both simple superpositions of two positive-momentum states and general superpositions thereof. For comparison, in Section VI, we undertake the “traditional” calculation of Bracken and Melloy  \cite{Bracken_1994, ring1}, which consists in the computation of the total probability flowing backwards at the origin during a time interval $T$, in order to estimate the constant $c_{\text{BM}}$. Finally, our conclusions are summarized in Section VII. 

\section{The models and their solutions}

 In this section we describe the two configurations of the tight-binding model under study, and their corresponding solutions. This shall also help to fix the notation followed throughout the paper.

%\subsection{The infinite tight-binding chain}
The tight-binding chain is described by the Hamiltonian (see fig.1):
\begin{equation}
    \mathbf{H}= -\tau\big((1+i\epsilon)\mathcal{\mathbf{S}}+ (1-i\epsilon)\mathcal{\mathbf{S}}^{\dagger}\big), 
    \label{eqn: hamiltonian_dis}
\end{equation}
where $\tau$ is a coupling strength, which sets the energy scale of the problem, and $\epsilon$ is a dimensionless parameter whose physical meaning will be discussed in Section \ref{IV}. We are considering constant onsite energies, so without loss of generality, we can set these to zero. The Hamiltonian in discrete configuration space is expressed  using the shift operators $\mathbf{S}$ and $\mathbf{S}^\dagger$. For a state in the site representation, $\psi_k(j)$, the actions of these operators are~\cite{Fukui, kelley2001difference} 
\begin{equation}
    \mathbf{S} \psi_k(j)=\psi_k(j+1), \quad  \mathbf{S}^{\dagger} \psi_k(j)= \psi_k(j-1).
\end{equation}
Here we observe that $\mathbf{H}\neq \mathbf{H}^*$, indicating the absence of time-reversal symmetry, however, as $\mathbf{H}=\mathbf{H}^\dagger$, the Hamiltonian remains Hermitian. The eigenstates  of the time-independent Schr{\"o}dinger equation
\begin{eqnarray}
    E_{k}\psi_{k}(j)=-\tau\big((1+i\epsilon)\psi_{k}(j+1)+ (1-i\epsilon)\psi_{k}(j-1)\big)\label{eqn: se},
\end{eqnarray}
are  plane waves
\begin{equation}
     \psi_{k}(j)= \frac{e^{ikj }}{\sqrt{2\pi}}.   
\end{equation}
The energy corresponding to the eigenfunction $ \psi_{k}(j)$ is 
\begin{equation}
E_k=-2 \tau \sqrt{1+\epsilon^2}\cos\left(k+\xi \right), \label{eq:energy}
\end{equation}
where $\xi= \arctan(\epsilon)$ and $k$ represents the pseudomomentum. For the purposes of this work, the pseudomomentum serves as a quantum number \cite{Streib} taking values in the interval $[-\pi, \pi]$ for the infinite chain. This quantity is not to be confused with the physical momentum operator $\mathbf{p}$ nor its eigenvalues, discussed in Sec.~\ref{sec:pop}. 
\begin{figure}
     \centering
     \includegraphics[width=0.7\linewidth]{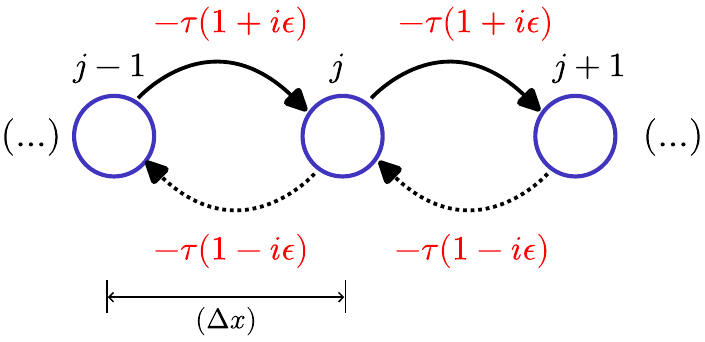}
     \caption{Segment of the chain, the lattice elements are separated by a distance $(\Delta x)$ and the nearest-neighbor hopping is characterized by a complex number with imaginary part $\pm\epsilon$.}
     \label{fig:placeholder}
 \end{figure}

When we impose periodic boundary conditions on the chain of $N$ sites ($\psi_{k}(0)=\psi_{k}(N)$), the pseudomomentum takes discrete values 
$$  k= \frac{2n\pi}{N},$$
where  $n$ is an integer ranging from $0$ to $N$. 

\section{Continuity equation}
\label{IV}
As mentioned in the previous section, we now provide a detailed discussion of the physical significance and implications of the "bias" parameter $\epsilon$. Let us consider a wave function $\Psi(j,t)$, solution to the time-dependent Schr\"odinger equation with Hamiltonian~(\ref{eqn: hamiltonian_dis}). The continuity equation for its probability density $|\Psi(j,t)|^2$ reads 
\begin{equation}
    \frac{\partial }{\partial  t} |\Psi(j,t)|^2= -J(j+1,t)+J(j,t),\label{eq:conteq}
\end{equation}
where $J(j,t)$ is the flux from the site $j-1$ to $j$, at time $t$, and is given by 
\begin{equation}
   J(j,t)= \frac{\tau }{i \hbar} \Big(\Psi^*(j-1,t)\Psi(j,t)-\Psi(j-1,t)\Psi^*(j,t)+i\epsilon\big(\Psi^*(j-1,t)\Psi(j,t)+\Psi(j-1,t)\Psi^*(j,t)\big)\Big).
   \label{eqn: flux1}
\end{equation}
Now, we consider the continuous limit of eq.(\ref{eq:conteq}), which can be obtained by recalling that the position $x$ is measured in terms of the fundamental length $(\Delta x)$ through the relation $x = (\Delta x) j$ \cite{weiss1994aspects}. Then, taking $(\Delta x)\rightarrow 0 $, we can expand the wave function in a Taylor series as
\begin{equation}
     \Psi((\Delta x) (j\pm1),t)\simeq \Psi(x,t)\pm(\Delta x) \frac{\partial}{\partial x} \Psi(x,t)+\frac{(\Delta x)^2}{2}\frac{\partial^2}{\partial x^2} \Psi(x,t)+\dots, 
\end{equation}
      
Using eqs.(\ref{eq:conteq}) and (\ref{eqn: flux1}), we obtain that the continuous expression for the flux is 
\begin{equation}
J(x,t)=  \frac{\hbar}{2\mu i} \Bigg[\Psi^*(x,t)\frac{\partial}{\partial x} \Psi(x,t)-\Psi(x,t)\frac{\partial}{\partial x} \Psi^*(x,t) +2i\zeta |\Psi(x,t)|^2 \Bigg], 
\label{eqn: cec}
\end{equation}
where the constants $\tau$ and $\epsilon$ are given by:
\begin{equation}
    \tau=\frac{\hbar^2}{2\mu (\Delta x)^2}, \qquad \epsilon=\zeta (\Delta  x), 
    \label{eqn: escv}
\end{equation}
consistently with the finite-difference approximation \cite{Timothy, Inui_2023}.

The first terms on the RHS of eq.(\ref{eqn: cec}) are the "standard" probability flux of a quantum particle, while the last can be identified as a "drift" term, much like what happens in biased diffusion \cite{weiss1994aspects}.  Applying the same approximations to the Hamiltonian in eq.(\ref{eqn: hamiltonian_dis}) yields the time-dependent Schr\"odinger equation
\begin{equation}
    i\hbar \frac{\partial }{\partial t} \Psi(x, t)=- \left(\frac{\hbar^2}{2\mu}\frac{\partial^2}{\partial x^2} +i \frac{\zeta \hbar ^2}{\mu} \frac{\partial}{\partial x} \right)\Psi(x,t) \label{eqn: se3},
\end{equation}
where we omitted an irrelevant constant energy term.

A physical realization of a quantum system with such a drift can be obtained, for example, by considering the Hamiltonian of a free particle confined to a ring in the $xy$ plane with radius $r_0$, centered at the origin, in the presence of a constant magnetic field of the form $\vec{B} = B_0 \hat{z}$ \cite{GVugalter_2004}. The Hamiltonian for a free particle in an external magnetic field reads
\begin{equation}
\mathbf{H} = \frac{1}{2\mu} \left(\mathbf{p} - e A \mathbb{I} \right)^2 ,
\end{equation}
where $e$ is the charge of the particle, $\mathbb{I}$ is the identity operator, and for this system, $\mathbf{p}$ can be written as
\begin{equation}
\mathbf{p} = \frac{p_{\vartheta}}{r_0} \, \hat{\vartheta} ,
\end{equation}
with $p_{\vartheta}$ being the momentum associated with the angular coordinate, which can be identified as the angular momentum.

For the magnetic field $\vec{B}$, the corresponding vector potential can be written as
\begin{equation}
\vec{A}  = -r_0 B_0 \hat{\vartheta} .
\end{equation}
Thus, the Hamiltonian becomes
\begin{equation}
\mathbf{H} = \frac{1}{2\mu} \left( \frac{\mathbf{L_z}}{r_0} + e r_0 B_0 \mathbb{I}\right)^2 ,
\label{eqn:circle}
\end{equation}
where the operator $\mathbf{L}_{z}$ is defined as usual by
$
\mathbf{L}_z = -i\hbar \frac{\partial}{\partial \vartheta} .
$

The term proportional to $B_0^2$, after expanding the square,  produces an irrelevant constant energy offset which can be ignored, so the Hamiltonian eq.(\ref{eqn:circle}) reduces to
\begin{equation}
\mathbf{H}' = -\frac{\hbar^2}{2\mu} \left( \frac{1}{r_0^2} \frac{\partial^2}{\partial \vartheta^2} + i\frac{eB_0}{\hbar} \frac{\partial}{\partial \vartheta} \right). 
\end{equation}
This is the Hamiltonian of a rigid rotor in a magnetic field \cite{GVugalter_2004}, which is analogous to the  Hamiltonian in equation (\ref{eqn: se3}), and can be regarded as the continuous counterpart of the Hamiltonian in eq. (\ref{eqn: hamiltonian_dis}) with periodic boundary conditions.

A different interpretation is reached recalling that the appearance of the drift term in diffusive systems is associated with a moving reference frame, in which case such term can be eliminated by means of a simple Galilean transformation. It is easy to see that with the transformation \cite{Rosen, Greenberger}
\begin{equation}
    x'=x-\frac{\hbar \zeta}{\mu}t, \qquad  t'=t.
\end{equation}
in the time-dependent Schr\"odinger equation eq.(\ref{eqn: se3}),  we get rid of the drift-like flux and recover the standard quantum flux.

The drift also has the effect of changing the range of pseudomomentum values $k$ that correspond to states with positive-momentum, as we show in the next section.

\section{The momentum operator in tight-binding systems}
\label{sec:pop}
 Before starting our analysis of quantum backflow in this system, it is necessary to determine the momentum operator $\mathbf{p}$ of a biased tight-binding chain in order to establish the values of the pseudomomentum 
$k$ for which its eigenvalues are non-negative. \\

We can write the momentum operator using the Heisenberg equation \cite{JTaguena, Gu}, 
\begin{equation}
   \mathbf{p}=\mu \frac{\partial \mathbf{r(t)}}{\partial t}= \frac{\mu}{i \hbar }[\mathbf{r}(t),\mathbf{H}], \label{Eq:Heis}
\end{equation}
where $\mu$ is the mass of the particle and  the position takes the values  $(\Delta x) j$, with $j$ integer (see fig.\ref{fig:placeholder})~\cite{Si_2025, Timothy} .

The time  evolution of the position operator can be obtained using the Baker–Campbell–Hausdorff lemma \cite{sakurai2017modern}:
\begin{equation}
  \mathbf{r}(t)= e^{i\mathbf{H}t/\hbar} \mathbf{r}e^{-i\mathbf{H}t/\hbar}=\mathbf{r}+[\mathbf{H},\mathbf{r}]\frac{i t}{\hbar}+[\mathbf{H}, [\mathbf{H},\mathbf{r}]]\frac{(it/\hbar)^2}{2! }+[\mathbf{H}, [\mathbf{H}, [\mathbf{H},\mathbf{r}]]]\frac{(it/\hbar)^3}{3! }+\dots .
    \label{eqn:p1}
\end{equation}

Computing the commutators yields
\begin{equation}
[\mathbf{H}, \mathbf{r}] = -(\Delta x)\tau\big((1+i\epsilon)\mathcal{\mathbf{S}}- (1-i\epsilon)\mathcal{\mathbf{S}}^{\dagger}\big)
\label{eqn: comm}
\end{equation}
and
\begin{equation}
      [\mathbf{H}, [\mathbf{H}, \mathbf{r}]]= 0.
      \label{eq:pcons}
\end{equation}
The last expression means that momentum is a conserved quantity and  shares eigenfunctions with the Hamiltonian. Using eqs.~(\ref{Eq:Heis}), (\ref{eqn:p1}) and (\ref{eq:pcons})  one obtains
\begin{equation}
    \mathbf{r}(t)= \mathbf{r}+\frac{\mathbf{p}}{\mu }t.
\end{equation}
As expected, this is the position operator for rectilinear motion, since the Hamiltonian is a discretization of a one-dimensional system where the particle moves freely. The momentum operator is
\begin{equation}
 \mathbf{p}=
    \frac{(\Delta x)\mu  \tau }{i \hbar}    \Big((1+i\epsilon)\mathbf{S}-(1-i\epsilon)\mathbf{S}^{\dagger}\Big), 
\end{equation}
and its eigenvalue equation is
\begin{equation}
    \mathbf{p} \psi_{k}(j)=  \frac{2(\Delta x)\mu\tau \sqrt{1+\epsilon^2}}{ \hbar} \sin\left( k +\xi\right) \psi_{k}(j). 
    \label{eq:peigeq}
\end{equation}
This agrees with the well-known relation between energy (c.f. eq.~(\ref{eq:energy})) and the momentum eigenvalue for tight-binding systems \cite{JTaguena, Pedersen, Lee2018, Boykin} 
\begin{equation}
\frac{(\Delta x)\mu }{\hbar}\frac{\partial E}{\partial k}=  \frac{2(\Delta x)\mu\tau \sqrt{1+\epsilon^2}}{ \hbar} \sin\left( k +\xi\right) .
\end{equation}
From eq.~(\ref{eq:peigeq}),  it can be seen that for the infinite chain, eigenstates with positive-momenta occur when the following condition 
\begin{equation}
 k \in \left[ -\xi,\; \pi-\xi \right]
\label{eqn: k_int}
\end{equation}
is satisfied. 

For the periodic chain; the eigenstates with positive-momentum will be those for which \begin{equation}
    n\in \left\{\eta_1 ,\eta_1 +1, \dots,\eta_2 -1, \eta_2 \right\}, \label{eqn: set1}
\end{equation}  where $\eta_1= \left\lceil -\dfrac{\xi}{2\pi}\,N \right\rceil $ and $\eta_2=\; \left\lfloor \dfrac{\pi-\xi}{2\pi}\,N \right\rfloor$. Here, $\lfloor x\rfloor$ and  $\lceil x\rceil $ denote the usual floor and ceil functions.  

\section{Maximal backflow and bounds on probability density flow}
\label{sec:max}
We start our analysis of quantum backflow by obtaining the states that maximize the amplitude of the negative flux. First, we illustrate the key idea using a wave function composed of two eigenstates, and then extend the approach to wave packets formed from all positive-momentum eigenstates in both the periodic and infinite cases. 
\subsection{An illustrative example: The two-state backflow}
We consider a wave function $\Psi(j,t)$ comprised of two eigenstates $\psi_{k(m_1)}(j,t)$ and $\psi_{k(m_2)}(j,t )$ with $m_1,m_2$ belonging to the set of eq.~(\ref{eqn: set1}) so $\Psi(j,t)$ is the simplest superposition of eigenstates that have positive-momentum:
\begin{equation}
  \Psi(j, t)=\frac{\cos\left(\frac{\theta}{2}\right)e^{i\left(\frac{2\pi m_1 j}{N}-\frac{E_{m_1}}{\hbar}t\right)}+\sin\left(\frac{\theta}{2}\right)e^{i\left(\frac{2\pi m_2 j}{N}-\frac{E_{m_2} }{\hbar}t+\gamma\right)}}{\sqrt{N}},
\end{equation}
where  $\theta\in [0, \pi]$ and $\gamma\in[0, 2\pi ]$.  %The density per site for this wave function is 

%\begin{equation}
%|\Psi(j, t)|^2=  \frac{1 + \cos\Big(\frac{ 2 j (m_1 - m_2) \pi }{ N } - \frac{E_{m_1} - E_{m_2} }{\hbar } t - \gamma \Big) \sin(\theta)}{N}.
%\end{equation}
The flux computed via eq. (\ref{eqn: flux1}) is 
\begin{equation}
\begin{split}
    J(j,t )  &= \scriptstyle \frac{2 \tau \sqrt{1+\epsilon^2} }{ N\hbar}  \Big[   \cos\left(\frac{(m_1 - m_2)\pi}{N}\right)  \sin\left(\frac{(m_1 + m_2)\pi}{N}  +\xi\right)  + 
 \cos(\theta) \sin\left(\frac{(m_1 - m_2)\pi}{N}\right)  \cos\left( \frac{(m_1 + m_2)\pi}{N} +\xi\right) 
\\&\scriptstyle  +\cos\left( \frac{(2j-1)(m_1 - m_2)\pi}{N} - \frac{E_{m_1}-E_{m_2} }{\hbar}t - \gamma \right)\sin\left( \frac{(m_1 + m_2)\pi}{N}+ \xi \right) 
    \sin(\theta)\Big], \label{eq:Jtwo}
\end{split}
\end{equation}
Note that the term $ \sin\left( \frac{(m_1 + m_2)\pi}{N}+\xi \right) $  always takes positive values for the current choice of $m_1$ and $m_2$. This can be seen by assuming, without loss of generality, that  $m_1<m_2$, then $m_1 \leq \frac{m_1+m_2}{2}\leq m_2$, so $\frac{(m_1+m_2)\pi}{N}$ corresponds to a point between the bounds of the set defined in eq. (\ref{eqn: set1}). Thus, the minimum of the flux for a given $j$ and $\gamma$ fixed, is reached at times $t$ such that the cosine factor in which $t$ appears is -1, that is
\begin{equation}
\begin{split}
\mathcal{J}(\theta)=  \scriptstyle    \min\limits_{\{t\}}(J(j,t ))   &=\scriptstyle\frac{2 \tau \sqrt{1+\epsilon^2} }{N \hbar}  \Big[    \cos\left(\frac{(m_1 - m_2)\pi}{N}\right)  \sin\left(\frac{(m_1 + m_2)\pi}{N}  +\xi\right)  + 
 \cos(\theta) \sin\left(\frac{(m_1 - m_2)\pi}{N}\right)  \cos\left( \frac{(m_1 + m_2)\pi}{N} +\xi\right) 
\\&\scriptstyle -
     \sin\left( \frac{(m_1 + m_2)\pi}{N}+ \xi \right) 
    \sin(\theta)
\Big].
\end{split}
\end{equation}
Now we can adjust the parameter $\theta$  to obtain the minimum flux given the two states with positive-momenta. First, we rewrite the last equation as 
\begin{equation}
    \mathcal{J}(\theta)=\frac{2 \tau \sqrt{1+\epsilon^2} }{N \hbar}(a+b\cos(\theta)-c\sin(\theta)) \label{eqn: minj}, 
\end{equation}
where 
\begin{align}
    a&=   \cos\left(\frac {(m_1 - m_2) \pi}{ N}\right)\sin\left(\frac {(m_1 + m_2) \pi}{ N}+\xi\right), \\\nonumber
b&=  \sin\left(\frac {(m_1 - m_2) \pi}{ N}\right)\cos\left(\frac {(m_1 + m_2) \pi}{ N}+\xi\right), \\\nonumber
    c&=  \sin\left(\frac {(m_1 + m_2) \pi}{ N}+\xi\right).
\end{align}
Minimizing the expression of eq. (\ref{eqn: minj}) with respect to $\theta$, we find that 
\begin{equation}
\tan(\theta)= -\frac{c}{b}, 
\end{equation}
 Thus, the global minimum of $J(j,t)$ is 
\begin{equation}
    \min_{\{t,\theta\}}(J(j,t))=\frac{2 \tau \sqrt{1+\epsilon^2} }{ N\hbar}(a-\sqrt{b^2+c^2}).
    \label{eqn: max_2}
\end{equation}
Note that $a <\sqrt{b^2+c^2}$ as long as $m_1\neq m_2$, so the lowest flux attainable from the superposition of plane waves with positive-momenta is, indeed, negative.

\subsection{Infinite chain}
We now address the general case, starting  with the infinite chain. In view of eq.~(\ref{eqn: k_int}), the most general wave function composed of states with positive-momentum is given by
\begin{equation}
    \psi(j,t)=\frac{1}{\sqrt{2\pi}}\int_{-\xi}^{\pi-\xi}  e^{i\left(kj-\frac{E_{k}}{\hbar}t\right)} \phi(k)\,dk.
    \label{eqn: wf1}
\end{equation}
  Here $\phi(k)$ is a weight function that satisfies the normalization condition 
\begin{equation}
   \int_{-\xi}^{\pi-\xi} |\phi(k)|^2\,dk=1.
   \label{eqn: norm}
\end{equation}

The flux associated with the wave function $\psi(j,t)$ in eq.~(\ref{eqn: wf1}) can be computed from  eq.(\ref{eqn: flux1}) as a functional of $\phi(k)$:
\begin{equation}
 J(j,t)= \frac{\tau\sqrt{1+\epsilon^2}}{\pi \hbar} \int_{-\xi}^{\pi-\xi}  \int_{-\xi}^{\pi-\xi} \phi^*(k')\phi(k) 
\, e^{ i\left( (j-\frac{1}{2} )(k - k') + \frac{(E_{k'} - E_{k}) t}{\hbar} \right)} 
 \sin\left( \frac{k + k'}{2} +\xi\right)  \,dk\,dk',
 \label{eqn: flux2}
\end{equation}
We want to minimize the flux at a given position $j'$ and time $t'$, subject to the constraint of eq. (\ref{eqn: norm}). To do this, we define the functional
\begin{equation}
\begin{split}
 \scriptscriptstyle I_1( \phi^*(k'),\phi(k) )= \frac{\tau \sqrt{1+\epsilon^2}}{\pi \hbar} \int_{-\xi}^{\pi-\xi}  \int_{-\xi}^{\pi-\xi} \phi^*(k')\phi(k) 
\, e^{ i\left( (j'-\frac{1}{2} )(k - k') + \frac{(E_{k'} - E_{k}) t'}{\hbar} \right)} 
 \sin\left( \frac{k + k'}{2} +\xi\right) \,dk\,dk' - \lambda \int_{-\xi}^{\pi-\xi} |\phi(k')|^2\,dk', 
\end{split}
\end{equation}
where $\lambda$ is a Lagrange multiplier. From this expression, the following Euler-Lagrange equation can be derived:
\begin{equation}
\begin{split}
     \frac{\tau\sqrt{1+\epsilon^2}}{\pi \hbar} \int_{-\xi}^{\pi-\xi} \phi(k) 
\, e^{ i\left( (j'-\frac{1}{2} )(k - k') + \frac{(E_{k'} - E_{k}) t'}{\hbar} \right)} \sin\left( \frac{k + k'}{2} +\xi\right)
 \,dk=\lambda\phi(k'),
 \end{split}
 \end{equation}
 or
 \begin{equation}
\scriptstyle\frac{\tau\sqrt{1+\epsilon^2}}{\pi \hbar}  \int_{-\xi}^{\pi-\xi} \phi(k) \,  e^{ i\left( (j'-\frac{1}{2} )(k - k') + \frac{(E_{k'} - E_{k}) t'}{\hbar} \right)} 
\, 
\left[\sin\left( \frac{k + \xi}{2} \right)\cos\left( \frac{k' + \xi}{2} \right)  +\cos\left( \frac{k + \xi}{2}\right)\sin\left( \frac{k' + \xi}{2} \right) \right]\,dk=\lambda\phi(k'), 
\label{eqn: ee1}
\end{equation}
The last equation is a homogeneous Fredholm integral equation that can be solved by direct computation \cite{wazwaz2011linear}. First we set 
\begin{align}
\int_{-\xi}^{\pi - \xi} \phi(k)\, e^{i(k(j'+\frac{1}{2})+  t'\tau \sqrt{1 + \epsilon^2} \cos(k + \xi))} 
\sin\left( \frac{k + \xi}{2} \right)\, dk &= \alpha, 
\label{alpha}
\\[5pt]
\int_{-\xi}^{\pi - \xi} \phi(k)\,e^{i(k(j'+\frac{1}{2})+  t'\tau \sqrt{1 + \epsilon^2} \cos(k + \xi))} 
\cos\left( \frac{k + \xi}{2} \right)\, dk &= \beta, 
\label{beta}
\end{align}
thus, 
\begin{equation}
\lambda \, \phi(k') = 
\, e^{-i(k'(j'+\frac{1}{2})+  t'\tau \sqrt{1 + \epsilon^2} \cos(k' + \xi))} 
\left[ 
\alpha \cos\left( \frac{k' + \xi}{2} \right) + 
\beta \sin\left( \frac{k' + \xi}{2} \right) 
\right].
\label{eq:hilambda}
\end{equation}
Substituting this expression in eqs.~(\ref{alpha}) and (\ref{beta}), and evaluating the integrals, gives the following eigenvalue equation
\begin{equation}
\frac{\sqrt{1+\epsilon^2}\tau}{\pi \hbar}
\begin{bmatrix}
1 & \quad\frac{\pi}{2}\\
\frac{\pi}{2} &\quad 1
\end{bmatrix}
\begin{bmatrix}
\alpha\\
\beta
\end{bmatrix}= \lambda 
\begin{bmatrix}
\alpha\\
\beta
\end{bmatrix}, 
\end{equation}
from which we can compute $\lambda$ 
\begin{equation}
   \lambda_{\pm}= \frac{\left( 2 \pm \pi \right) \sqrt{1 + \epsilon^{2}} \, \tau}{2 \hbar \pi}, 
\end{equation}
then, we have that 
\begin{equation}
    \alpha=\pm\beta.
\end{equation}
We normalize the weight function $\phi(k')$ in eq.~(\ref{eq:hilambda}) by choosing
\begin{equation}
    \beta=\frac{1}{\sqrt{\pi\pm 2}}, 
\end{equation}
so we find 
\begin{equation}
   \phi_{\pm}(k')=  \frac{e^{-i(k'(j'+\frac{1}{2})+ \frac{t'\tau}{\hbar}\sqrt{1 + \epsilon^2} \cos(k' + \xi))}  }{\sqrt{\pi\pm 2}}\left[\pm \cos\left( \frac{ k'+\xi}{2} \right)+ \sin\left( \frac{ k'+\xi}{2} \right) \right].
   \label{eqn: phi1}
\end{equation}
The flux is then given by 
\begin{equation}\hskip -2cm 
\begin{split}
    J(j;j',t;t')&= \frac{\tau \sqrt{1+\epsilon^2}}{\pi (\pi\pm 2)\hbar} \int_{-\xi}^{\pi-\xi}  \int_{-\xi}^{\pi-\xi} 
\, e^{i((k-k')(j-j')+  \sqrt{1 + \epsilon^2}\tau(t-t')(  \cos(k + \xi))- \cos(k' + \xi)))} 
\\& \scriptstyle \left(\pm \cos\left( \frac{ k'+\xi}{2} \right)+ \sin\left( \frac{ k'+\xi}{2} \right) \right) \left(\pm \cos\left( \frac{ k+\xi}{2} \right)+ \sin\left( \frac{ k+\xi}{2} \right) \right) \sin\left( \frac{k + k'}{2}+\xi \right)  \,dk\,dk'.
\label{eqn: flux_k}
\end{split}
\end{equation}
In fig.~2a, we illustrate the probability density flux given by eq.~(\ref{eqn: flux_k}) as a function of time for position $j = j'=3$, and its vicinity. We take take  $t'=3$, so the extreme values of the flux occur at site $j=3$ at time $t=3$ as observed. We also show the bounds associated with the corresponding eigenvalues $\lambda_\pm$. The insets show that the largest (smallest) flux values indeed coincide with the upper (lower) bounds at the correct positions and times, as expected.

\subsection{Periodic chain}
The procedure developed for the infinite chain can now be adapted and applied to the periodic-chain case, taking into account, of course, the fact that the spectrum is discrete. We begin by noting that, according to eq.~(\ref{eqn: set1}), the most general superposition of eigenstates with non-negative momentum can be written as
\begin{equation}
    \Psi(j,t)=\sum_{n=\eta_1}^{\eta_2}  c_{n} a_{n}(j,t)= \frac{1}{\sqrt{N}}\sum_{n=\eta_1}^{\eta_2} e^{i\left(\frac{2 \pi n j}{N}-E_n t\right)} c_{n} 
    \label{eqn:wf2}, 
\end{equation}
recalling that $\eta_2= \lfloor \frac{\pi-\xi}{2\pi} N\rfloor$ and $\eta_1= \lceil -\frac{\xi}{2\pi}N \rceil$.  The coefficients $c_{m}$ satisfy the normalization condition
\begin{equation}
    \sum_{n=\eta_1}^{\eta_2 } |c_{n}|^{2}=1. 
    \label{eqn: cond1}
\end{equation}
The flux can be found using eq.~(\ref{eqn: flux1}), i.e.
\begin{eqnarray}
    J(j,t)=\frac{2 \tau\sqrt{1+\epsilon^2} }{N \hbar}\sum_{n=\eta_1}^{\eta_2} \sum_{m=\eta_1}^{\eta_2}c^*_{m}c_{n} \, e^{i\left( \frac{(2j-1)(m-n)\pi}{N} -\frac{(E_n - E_m) t}{\hbar} \right)} 
 \sin\left( \frac{(m + n)\pi}{N} +\xi \right).
 \label{eqn: dflux}
\end{eqnarray}
The derivation of the bounds is analogous to the infinite-chain case; however, instead of finding a function, we seek to determine the sequence $c_{m}$ that extremizes eq.~(\ref{eqn: dflux}) (see the Appendix for details).  Here again, we have $\max(J(j,t))=\lambda_{+}$ and $\min(J(j,t))=\lambda_{-}$ where
\begin{equation}
\begin{split}
     \lambda_{\pm}&=  \frac{\tau \sqrt{1+\epsilon^2}}{N \hbar } \Big[ \csc\!\left( \frac{\pi}{N} \right) 
\sin\!\left( \frac{\pi (\eta_2+1-\eta_1)}{N} \right) 
\sin\!\left( \frac{\pi (\eta_1 + \eta_2)}{N} + \xi \right)\\ &
\pm \sqrt{(\eta_2+1-\eta_1)^2- 
\left(\cos\!\left( \frac{\pi (\eta_1 + \eta_2)}{N} + \xi \right) 
\csc\!\left( \frac{\pi}{N} \right) 
\sin\!\left( \frac{\pi (\eta_2+1-\eta_1)}{N} \right)\right)^2}\Big].
\label{eqn: bound1}
\end{split}
\end{equation}
 The corresponding sequences associated with $\lambda_{\pm}$ are%
\begin{equation}
\begin{split}
     c_{m}^{\pm}=     \scriptstyle  \mathcal{N_{\pm}}   e^{-i\left( \frac{(2j'-1)m\pi}{N} -\frac{ E_mt'}{\hbar} \right)}\Big[\pm \sqrt{\eta_2+1-\eta_1-
\cos\!\left( \frac{\pi (\eta_1 + \eta_2)}{N} + \xi \right) 
\csc\!\left( \frac{\pi}{N} \right) 
\sin\!\left( \frac{\pi (\eta_2+1-\eta_1)}{N} \right)}  \cos\left(\frac{m}{N}\pi+\frac{\xi}{2}\right) \\ \scriptstyle  +\sqrt{\eta_2+1-\eta_1+
\cos\!\left( \frac{\pi (\eta_1 + \eta_2)}{N} + \xi \right) 
\csc\!\left( \frac{\pi}{N} \right) 
\sin\!\left( \frac{\pi (\eta_2+1-\eta_1)}{N} \right)} \sin\left(\frac{m}{N}\pi+\frac{\xi}{2}\right) \Big],
    \label{eqn: sec2}
\end{split}
\end{equation}
where  $\mathcal{N_{\pm}}$ is a normalization constant given by $\mathcal{N}_{\pm}=\frac{1}{\sqrt{\mathcal{D_{\pm}}}}$ with
\begin{flalign*}
\scriptstyle  \mathcal{D_{\pm}}= & (\eta_2+1-\eta_1)^{2} 
- \cos^{2}\!\left( \tfrac{\pi(\eta_{1}+\eta_{2})}{N} + \xi \right) 
\csc^{2}\!\left( \tfrac{\pi}{N} \right) 
\sin^{2}\!\left( \tfrac{\pi(\eta_2+1-\eta_1)}{N} \right) 
\\ &\scriptscriptstyle\pm\csc\!\left( \tfrac{\pi}{N} \right) 
\sin\!\left( \tfrac{\pi(\eta_2+1-\eta_1)}{N} \right) 
\sqrt{(\eta_2+1-\eta_1)^{2} - 
\cos^{2}\!\left( \tfrac{\pi(\eta_{1}+\eta_{2})}{N} + \xi \right) 
\csc^{2}\!\left( \tfrac{\pi}{N} \right) 
\sin^{2}\!\left( \tfrac{\pi(\eta_2+1-\eta_1)}{N} \right)} 
\sin\!\left( \tfrac{\pi(\eta_{1}+\eta_{2})}{N} + \xi \right). 
\end{flalign*}
In fig.~2b, we show the fluxes computed through eq.~(\ref{eqn: dflux}) for the sequence defined in eq.~(\ref{eqn: sec2}), together with their respective bounds $\lambda_{\pm}$, using the same values of $j'$ and $t'$ as in the infinite chain. The main difference with respect to that case lies in the fact that $J(j; j', t; t')$ does not rapidly decay to zero as $t$ increases; instead, it oscillates with smaller amplitudes, exhibiting intervals of attenuation and amplification, although it does not appear to reach the same magnitude as at $j'$ and $t'$.
\\

The case $\epsilon = 0$ simplifies considerably. In this situation, $\eta_1 = 0$ and we take $\eta_2$  as follows
$$\eta_2=\begin{cases}
   \frac{N}{2}-1, & \text{for }  N \text{ even} \\ \frac{N-1}{2}, & \text{for }  N \text{ odd}
\end{cases}$$ 
The choice of $\eta_2$ for odd $N$ is made in order to avoid the zero-momentum state being considered twice in the wave function defined in eq.(\ref{eqn:wf2}). Therefore, the bounds of the flux given in eq.(\ref{eqn: bound1}) become 
\begin{equation}
    \lambda_{\pm}=\begin{cases}
     \frac{\tau }{N \hbar } \left(\cot\left(\frac{\pi}{N}\right)\pm \sqrt{\frac{N^2}{4}-1}\right), &\text{for }  N \text{ even} \\ 
        \frac{\tau }{2N \hbar } \left(\cot\left(\frac{\pi}{2N}\right)\pm \sqrt{N(N+2)}\right), &\text{for }  N \text{ odd}
    \end{cases}
\end{equation}
 Thus,  we have that eq. (\ref{eqn: sec2})  reduces to
\begin{equation}
   \hskip -1cm c_{m}^{\pm}=\begin{cases}
        \frac{e^{-i\left( \frac{(2j'-1)m\pi}{N} -\frac{ E_m t'}{\hbar} \right)}}{\sqrt{\frac{1}{2}\left[N^2-2\pm2\sqrt{N^2-4}\cot\left(\frac{\pi}{N}\right)\right]}}\Big[ \pm \sqrt{N-2} \cos\left(\frac{m}{N}\pi\right) + \sqrt{N+2}\sin\left(\frac{m}{N}\pi\right) \Big], 
       & \text{for }  N \text{ even} \\ \\ 
        %\scriptstyle 
        \frac{e^{-i\left( \frac{(2j'-1)m\pi}{N} -\frac{ E_m t'}{\hbar} \right)}}{\sqrt{\frac{1}{2}\left[(N+1)^2\pm\sqrt{N(N+2)}\cot\left(\frac{\pi}{2N}\right)\right]}}\Big[ \pm  \sqrt{N}\cos\left(\frac{m}{N}\pi\right) + \sqrt{N+2}\sin\left(\frac{m}{N}\pi\right) \Big], & \text{for }  N \text{ odd}.
    \end{cases} 
\end{equation}

\begin{figure}[h]
    \centering
    \includegraphics[width=0.98\linewidth]{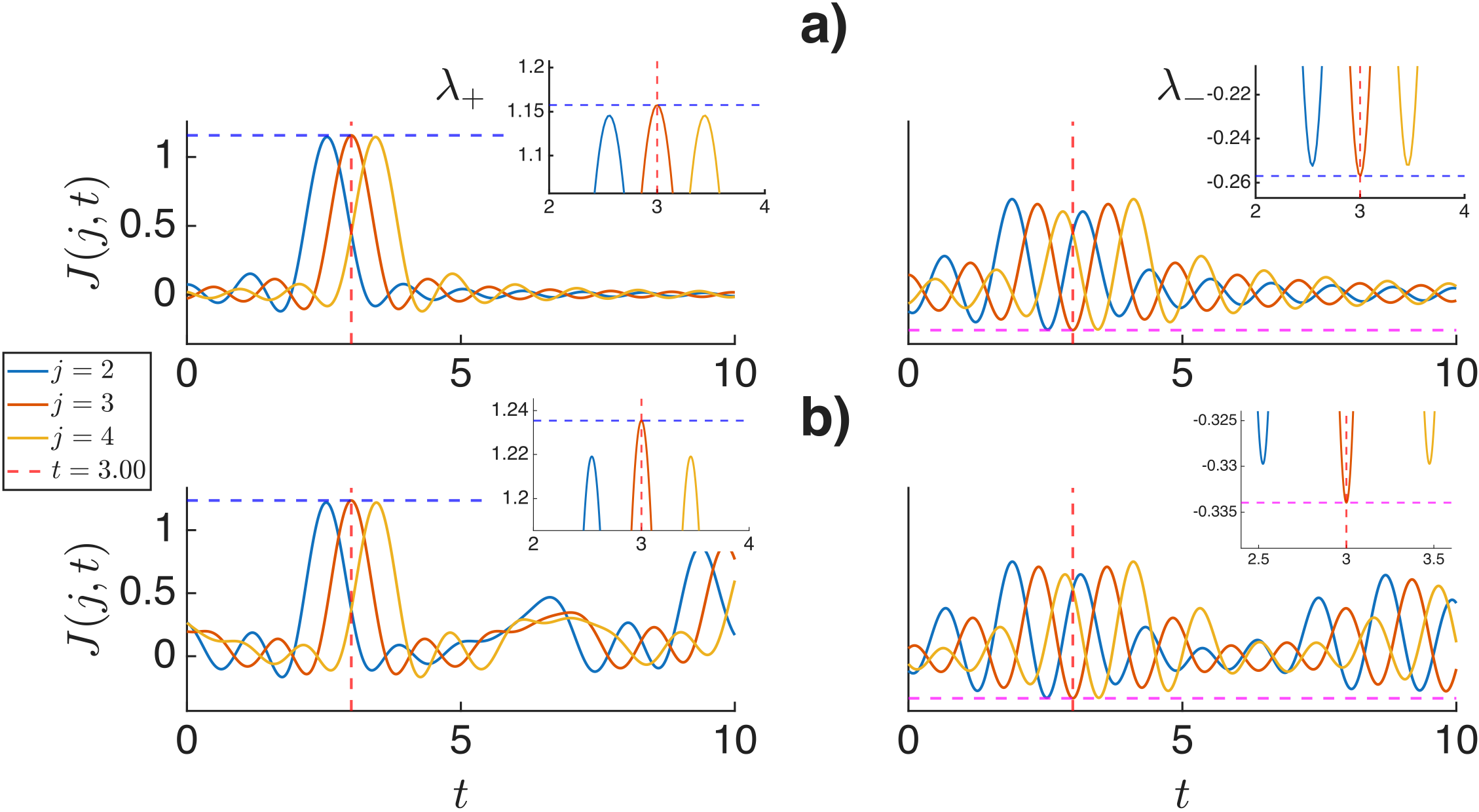}
    \caption{  The extreme values of the probability density flux at  $j' = 3$, and the nearest sites  are shown as a function of $t$ taking $t' = 3$ (dashed red line),  for the infinite chain (a) and the periodic chain (b) with $\epsilon=1$ and $N=9$. We compute the flux using the function $\phi_{\pm}(k)$ (sequence $c_m^{\pm}$) and the bound associated : $\lambda_{+}$ (blue dashed line) and $\lambda_{-}$ (blue dashed line). }
    \label{fig:fluxk}
\end{figure}

\begin{figure}[h]
    \centering
        \includegraphics[width=0.75\linewidth]{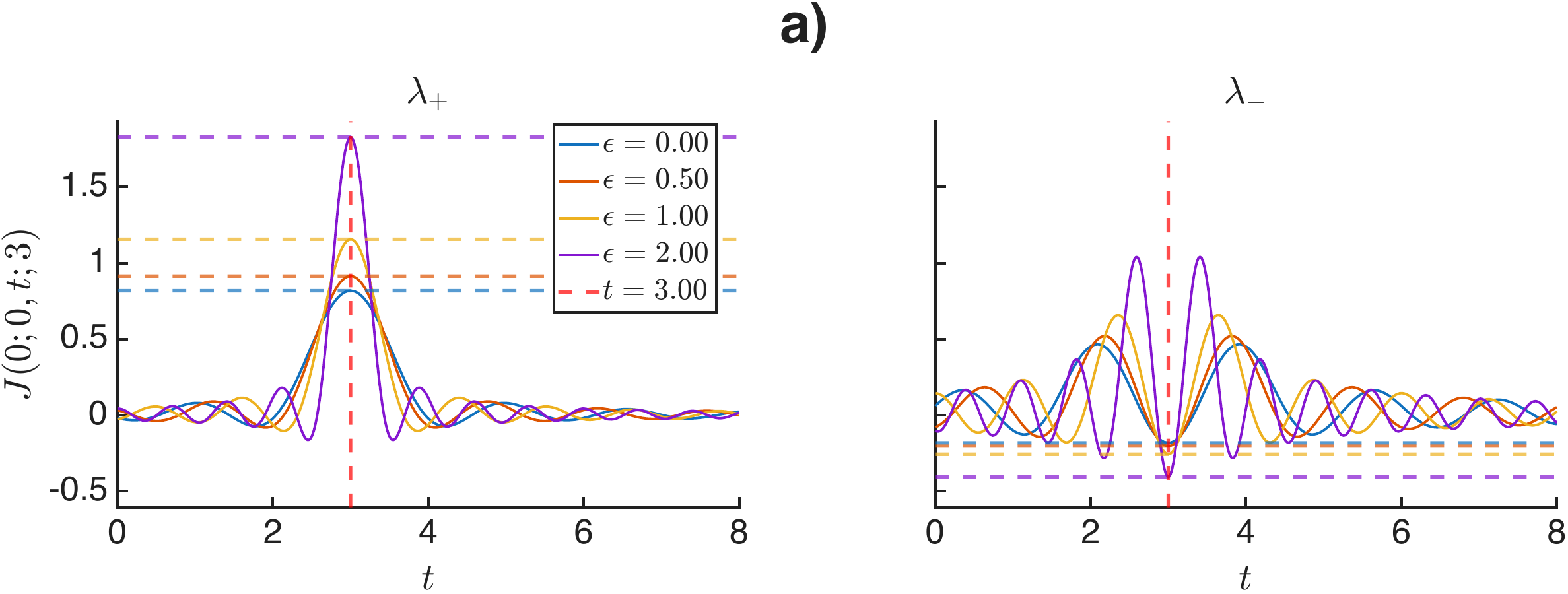}
         \includegraphics[width=0.75\linewidth]{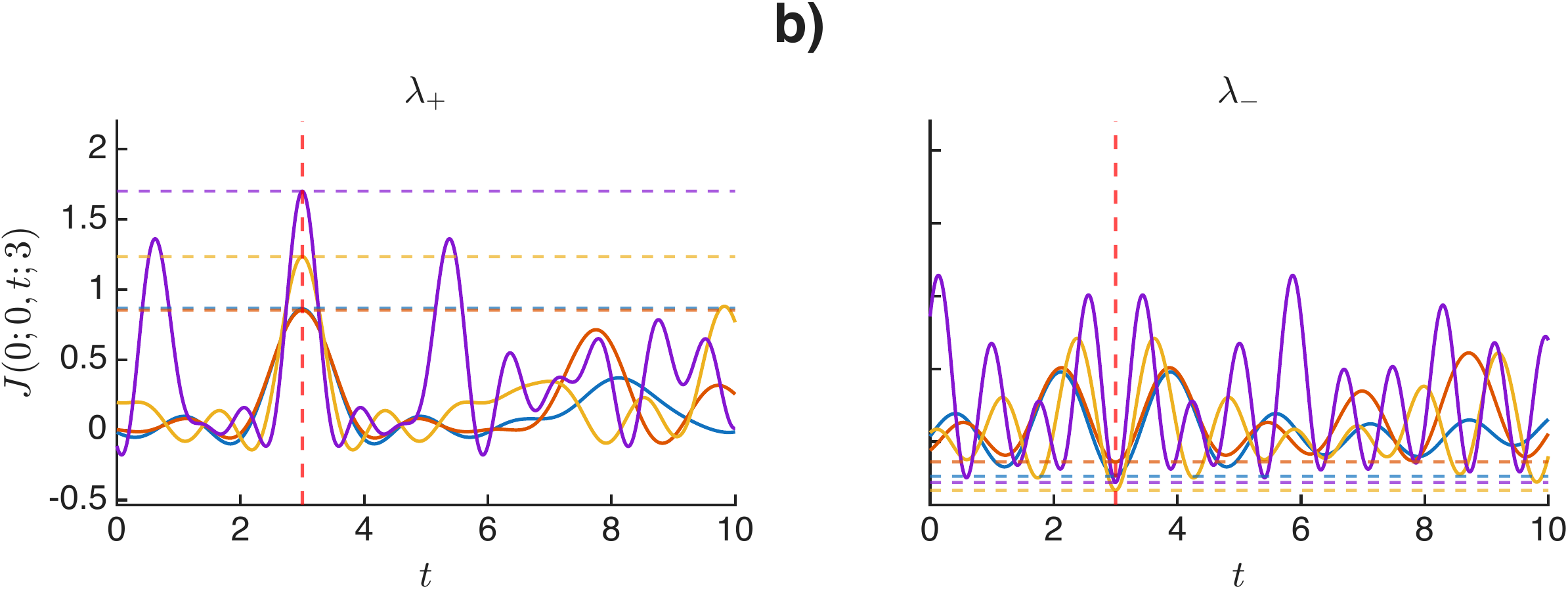}
    \caption{ The extreme values of the probability density flux for different $\epsilon$ values at  $j' = 0$, and $t' = 3$ (dashed red line) as a functions of time,  for the infinite chain (a) and the periodic chain with $N=9$ (b). The dashed lines are the respective bounds, $\lambda_{+}$  and $\lambda_{-}$ for the corresponding values of $\epsilon$.}
    \label{fig:fluxk2}
\end{figure}

The  effect  of the parameter $\epsilon$ on the probability density flux is shown in fig.~(\ref{fig:fluxk2}). We observe that the magnitude of the bounds on the flux increases with  $\epsilon$. This increase is more pronounced with $\lambda_{+}$, while it is comparatively milder with $\lambda_{-}$. An additional effect of the bias can be observed for the periodic chain, which exhibits larger oscillations of the flux far from the vicinity of $t=t'$, and these oscillations also increase with $\epsilon$.
\section{The Bracken-Melloy approach  to backflow}
Following the work of Bracken and Melloy  \cite{Bracken_1994}, we now focus on the amount of probability flowing backwards during a time interval  in our discrete tight-binding system. This description complements the analysis of the previous section. The original formulation of the problem aimed at estimating the maximum probability for a particle to be transferred from right to left across a given point (typically the origin) within a time interval $T$, for a wave function composed of a superposition of positive-momentum states in an infinite one-dimensional continuous system. This quantity is known as the Bracken--Melloy constant and is defined as
\begin{equation}
c_{\text{BM}} = \max \big(P(T) - P(0)\big),
\end{equation}
where $P(T)$ denotes the probability of finding the particle in the region $x < 0$, after a time $T$~\cite{Bracken_1994}. This is equivalent to determining the upper bound of the time integral of the probability current over a time window $[-T/2,\, T/2]$ at the origin, namely
\begin{equation}
c_{\text{BM}} = \max\!\left(-\int_{-T/2}^{T/2} J(0,t)\, dt\right).
\end{equation}
It is important to point out that,  for systems with periodic boundary conditions,  only the second definition of $c_{\text{BM}}$ makes sense.

As an extension of Bracken and Melloy's work, it has been shown that for continuous periodic systems, the total probability that flows to the left at the origin is bounded by 
$c^{\text{cont}}_{\text{ring}}\approx  3.0380 c_{\text{BM}}$~\cite{ring1}.

\subsection{Infinite chain} 

In this section we compute the extrema of the integrated flux for the case of our discrete infinite chain. First we  write the flux in eq.~(\ref{eqn: flux2}) at the origin as   
\begin{equation}
  J(0,t) =\frac{\tau \sqrt{1+\epsilon^2}}{\pi \hbar} \int_{-\xi}^{\pi-\xi}\int_{-\xi}^{\pi-\xi} \Phi^*(k')\Phi(k) 
    \, e^{ i\left(  \frac{(E_{k'} - E_{k}) t}{\hbar} \right)} 
 \sin\left( \frac{k + k'}{2}+\xi \right)  \,dk\,dk', 
\end{equation}
where $\Phi(k)=e^{-i\frac{k}{2}} \phi(k)$. 
Integrating over a temporal window $[-T/2, T/2]$ we obtain
\begin{equation}
\hskip -1cm   \int_{-T/2}^{T/2 } J(0, t)dt=  \int_{-\xi}^{\pi-\xi} \int_{-\xi}^{\pi-\xi} \frac{
  \sin\left( 2\nu\sqrt{1+\epsilon^2}\sin\left(\frac{k+k'}{2}+\xi \right) \sin\left(\frac{k-k'}{2}\right)\right)
}{2\pi \sin\left(\frac{k-k'}{2}\right)}\, \Phi^*(k')\Phi(k)dk dk', 
\end{equation}
where we introduced the dimensionless parameter $\nu=\frac{\tau T}{\hbar}$. 
In order to maximize this quantity subject to the constraint given by eq.~\eqref{eqn: norm}, we consider the functional:
\begin{equation}
\begin{split}
 \hskip -1cm     I_2(\Phi(k), \Phi^*(k')) =- \int_{-\xi}^{\pi-\xi} \int_{-\xi}^{\pi-\xi} \Phi^*(k')\Phi(k)
  \frac{
  \sin\left(2\nu\sqrt{1+\epsilon^2}\sin\left(\frac{k+k'}{2}+\xi \right) \sin\left(\frac{k-k'}{2}\right)\right)
}{2\pi \sin\left(\frac{k-k'}{2}\right)}\,dk\,dk'
\\ -\lambda_p \int_{-\xi}^{\pi-\xi} |\Phi(k')|^2 \,dk', 
\end{split}
\end{equation}
where $\lambda_p$ is a Lagrange multiplier.  Thus, we require the solution to the  Euler–Lagrange equation:
    \begin{equation}
\frac{1}{\pi}  \int_{-\xi}^{\pi-\xi} \Phi(k)
K(k,k')\,dk 
= \lambda_p\,\Phi(k'), 
\label{eqn: eig1}
\end{equation}
where the kernel $K(k,k')$ is given by
\begin{equation}
  K(k,k')= - \frac{
  \sin\left( 2\nu\sqrt{1+\epsilon^2}\sin\left(\frac{k+k'}{2}+\xi \right) \sin\left(\frac{k-k'}{2}\right) \right)
}{2\sin\left(\frac{k-k'}{2}\right)}, 
\end{equation}
 The continuous counterpart of the eigenvalue problem in eq.~(\ref{eqn: eig1}) can be obtained through the following rescaling relations: $k = (\Delta x)p$ and $k' = (\Delta x)q$. This leads to $\sin((\Delta x)p) \sim p(\Delta x) + o((\Delta x)^2)$. Writing $\tau$ as in eq. (\ref{eqn: escv}), we have $\nu = \frac{\hbar T}{2\mu (\Delta x)^2}$. Taking the limit $(\Delta x)\rightarrow 0$ we obtain

 \begin{equation}
   - \frac{1}{\pi}  \int_{-\zeta }^{\infty} \tilde{\Phi}(p)
\frac{\sin\left(\frac{T}{4 \mu\hbar }(p+q+\zeta)(p-q)\right)}{p-q}\,dp 
= \lambda\,\tilde{\Phi}(q) , 
\label{eqn: eig3}
 \end{equation}
 where $\tilde{\Phi}(p)= \lim_{(\Delta x)\rightarrow 0}\Phi((\Delta x) p)$ and $\zeta$ is defined in eq.~(\ref{eqn: escv}). 
 This is the "biased" version of the eigenvalue problem obtained by Bracken and Melloy \cite{Bracken_1994}.

Note that the kernel is the same  as the one obtained for a backflow problem under a constant force \cite{Melloy1, Goussev1} with the  difference that the lower integration limit in eq.~(\ref{eqn: eig3}) is shifted due to the effect of the drift. 

\begin{figure}[H]
        \centering
        \includegraphics[width=0.9\linewidth]{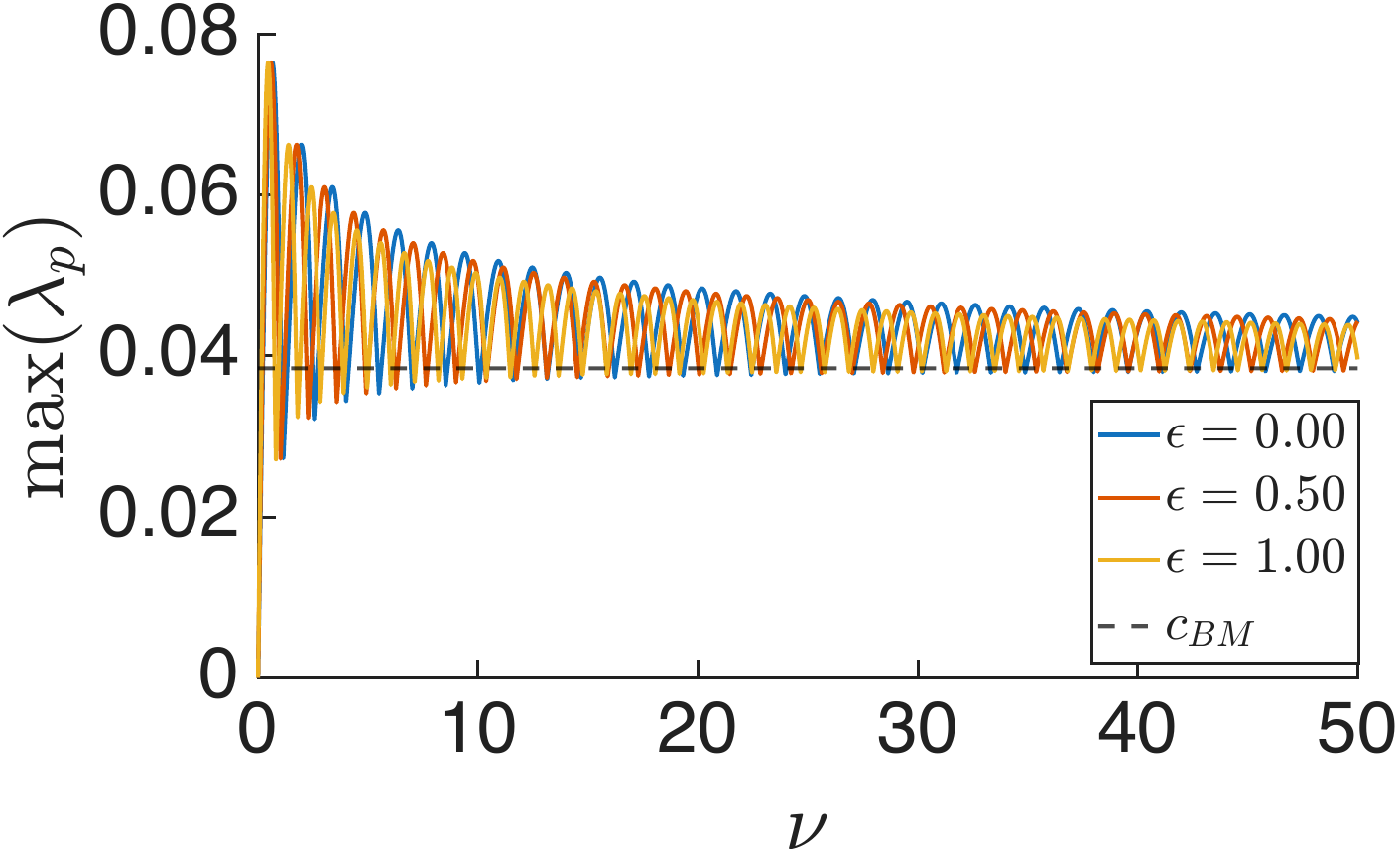}
        \caption{Supremum of the minus time-integrated probability flux, $\min(\lambda_p)$, as a function of $\nu$ for different values of $\epsilon$.
It can be observed that $\lambda_p$ appears to become bounded below by $c_{\text{BM}}$ as $\nu \rightarrow \infty$. The maximum of $\lambda_p$ tends to be larger for values of $\nu$ close to zero, and the effect of $\epsilon$ does not significantly change the qualitative behavior of the curves. We find that the three curves reach a maximum value of $ 0.0764734$.}
        \label{fig:4}
    \end{figure}

As occurs with its continuous counterpart,  the eigenvalue problem of eq. (\ref{eqn: eig1}) cannot be computed analytically, so we approach the problem numerically. We base the calculation on the fact that the integration can be approximated as a symmetric finite square matrix acting on a vector. The largest eigenvalue of the matrix is taken as an approximation to $\lambda_p$ \cite{Bracken_1994, eveson, Penz_2006}. 
Fig. (\ref{fig:4}) shows the dependence of the largest eigenvalue on the parameter $\nu$ for several values of $\epsilon$. The value of $c_{\text{BM}}$, which bounds the continuous case, is indicated in the figure. We observe that as $\nu$ increases, $\lambda_p$ appears to become bounded from below by $c_{\text{BM}}$. Here, $\epsilon$ appears to act only as some kind of scaling factor for each curve,  narrowing it towards the $y-$axis without actually affecting qualitatively the behavior of the curves.

\subsection{Periodic chain} 

We now turn our attention to the periodic chain. Using the expressions derived in Section~\ref{sec:max},   the integral of eq.~(\ref{eqn: dflux}) at $j=0$ over the temporal window $[-T/2, T/2]$ is:
\begin{equation}
%\begin{split}
\hskip -1cm \int_{-T/2}^{T/2} J(0,t)\, dt  \\%=
= \sum_{n=\eta_1}^{\eta_2} \sum_{m=\eta_1}^{\eta_2} C^*_{m} C_{n} 
\frac{\sin\left(
   2\nu  \sqrt{1+\epsilon^2}\sin\left(\frac{ (m+n) \pi}{N}+\xi \right)\sin\left(\frac{ (m-n) \pi}{N}\right)
  \right)
}{
 N \sin\left(\frac{(m - n)\pi}{N}\right)   
}.
%\end{split}
\end{equation}
where $    C_n=e^{-\frac{in\pi}{N}} c_n$. The extremes of this integral subject to the constraint eq.~(\ref{eqn: cond1}) can be found through the functional, 
\begin{equation}
\scriptstyle I_2
(C_m^*, C_n)=-\sum_{n=\eta_1}^{\eta_2} \sum_{m=\eta_1}^{\eta_2}C^*_{m}C_{n} \frac{
   e^{\frac{i (m- n) \pi}{N}} \, 
  \, 
  \sin\left( 2\nu  \sqrt{1+\epsilon^2}\sin\left(\frac{ (m+n) \pi}{N}+\xi \right)\sin\left(\frac{ (m-n) \pi}{N}\right)\right)  
}{
 N \sin\left(\frac{ (m-n) \pi}{N}\right)   
} -\lambda_p\sum_{m=\eta_1}^{\eta_2} |C_m|^2.
\label{eqn: f2}
\end{equation}

The Euler-Lagrange equation associated with eq. (\ref{eqn: f2}) leads to the eigenvalue equation 
\begin{equation}
     \sum_{n=\eta_1}^{\eta_2}  K_{m,n}  C_{n} = \lambda_p C_{m}.
     \label{eqn: eigen2}
\end{equation}
with
\begin{equation}
K_{m,n} = 
-\frac{1}{N}\begin{cases}

\frac{
   \, 
  \sin\left( 2\nu  \sqrt{1+\epsilon^2}\sin\left(\frac{ (m+n) \pi}{N}+\xi \right)\sin\left(\frac{ (m-n) \pi}{N}\right)\right)  
}{
  \sin\left(\frac{ (m-n) \pi}{N}\right)   
}
& \text{if } m \neq n,  \\[1.5ex]
   2\nu \sqrt{1+\epsilon^2}\sin\left( \frac{2n\pi}{N}+\xi \right) 
& \text{if } m = n, 
\end{cases}
\end{equation}
where the indices $m,n$ run from $\eta_1$ to $\eta_2$. This eigenvalue equation~(\ref{eqn: eigen2}) can be connected to its continuous counterpart \cite{ring1} using
\begin{equation}
    N= \frac{L}{(\Delta x)}
    \label{eqn: er}
\end{equation} 
and taking the limit $(\Delta x) \rightarrow 0$. \\

\begin{figure}[H]
    \includegraphics[width=0.95\linewidth]{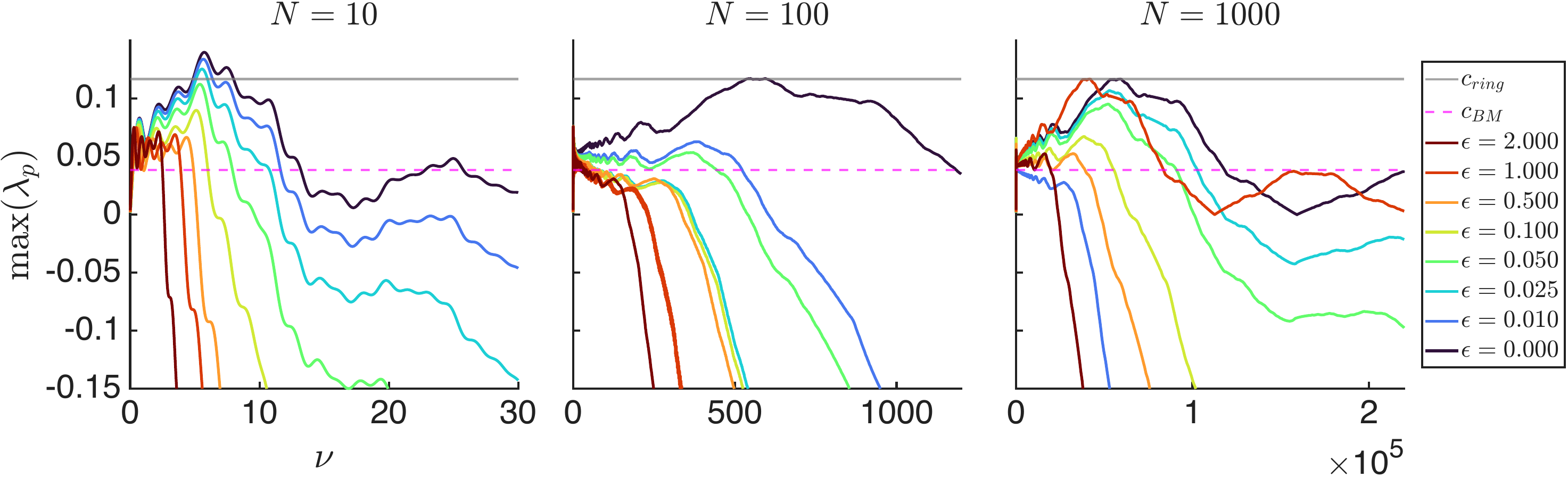}
    \caption{ Maximum of the time negative integrated probability density flux as a function of $\nu$, for $N=10, 10^2$ and  $10^3$ sites. The solid gray line corresponds to the bound for the continuous version of the system, while the blue dotted line denotes $c_{\text{BM}}$. We note that as the number of sites increases, the peak value of the curves converges toward $c^{\text{cont}}_{\text{ring}}$.}
    \label{fig:brakcen-melloy2}
\end{figure}

We solve numerically the eigenvalue problem of eq.~(\ref{eqn: eigen2}) in the same way as we did for the infinite chain case. 
Our results are shown in figure (\ref{fig:brakcen-melloy2}). There we display the maximum eigenvalue $\lambda_p$ for several values of $\epsilon$, highlighting $c_{\text{BM}}$ and $c^{\text{cont}}_{\text{ring}}$, increasing the size of the lattice by an order of magnitude in each figure.  We note that as the number of sites increases, the figures increasingly resemble those reported for the continuous ring by Goussev \cite{ring1}. Specifically, for $\epsilon = 0$, the curves display a shape to those obtained for the continuous ring, they have similar features and their maxima tend to $c^{\text{cont}}_{\text{ring}}$, corresponding to the gray line, as $N$ increases.\\

\begin{figure}[H]
    \centering
    \includegraphics[width=0.75\linewidth]{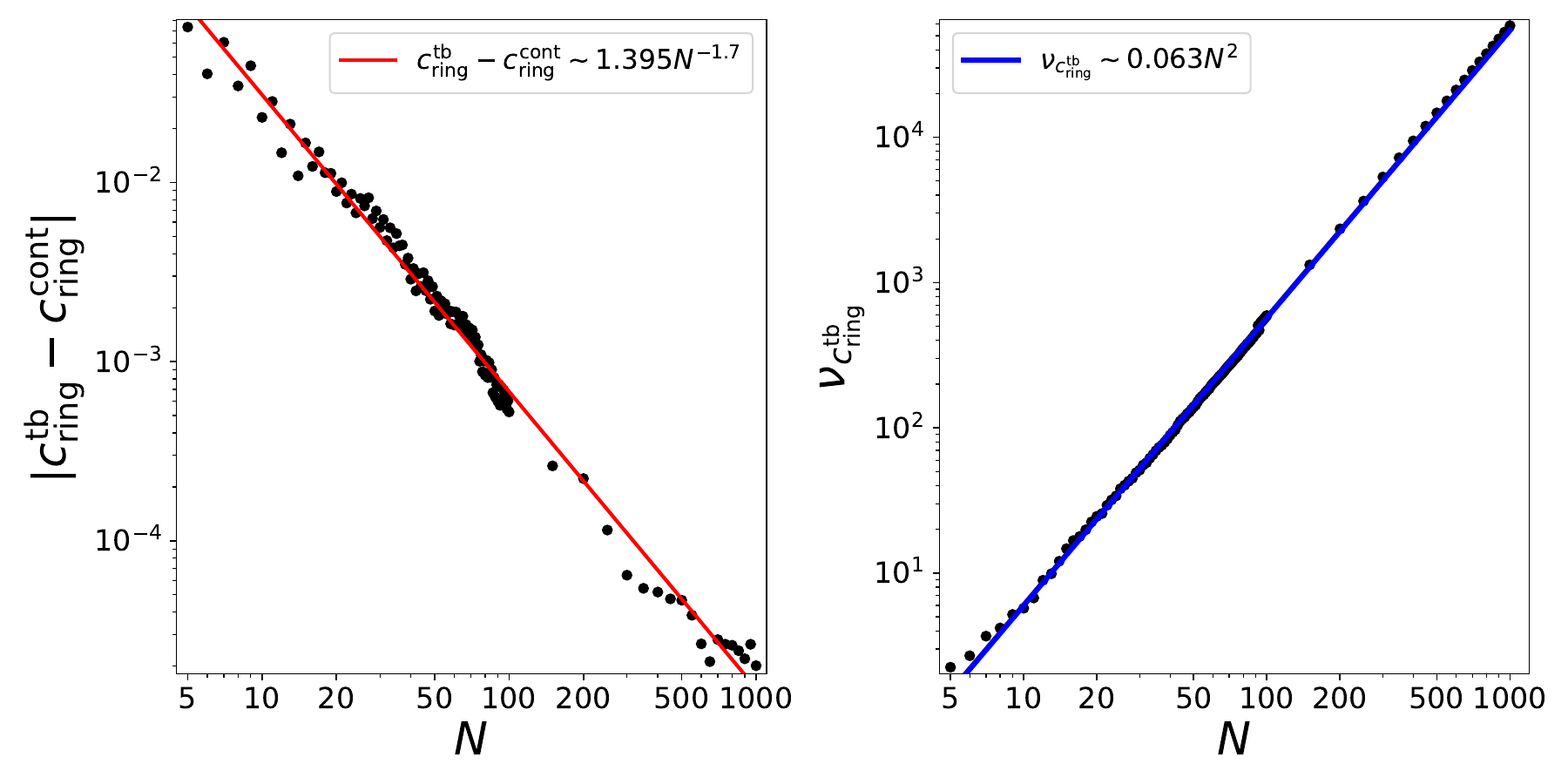}
    \caption{In the left panel, we show $c_{\text{ring}}^{\text{tb}} - c_{\text{ring}}^{\text{cont}}$, and in the right panel the value of $\nu_{c_{\text{ring}}^{\text{tb}}}$, both as a functions of $N$. Both curves appear to be well described by a power law. We include the best power law fits for each curve.}
    \label{fig: lambda_vs_N}
\end{figure}

Finally, for each given $N$, we denote by $c_{\text{ring}}^{\text{tb}}$ the maximum of the curve $\max(\lambda_p)$ as a function of $\nu$ when $\epsilon = 0$. In fig. (\ref{fig: lambda_vs_N}) we show how $c_{\text{ring}}^{\text{tb}}$  and its corresponding value of $\nu_{c_{\text{ring}}^{\text{tb}}}$ depend on $N$. To establish a clear relationship, we subtract from $c_{\text{ring}}^{\text{tb}}$ the value of $c^{\text{cont}}_{\text{ring}}=0.11681564947322964$ \cite{ring1} to which it tends asymptotically as $N^{-1.7}$ approximately. On the other hand, the right panel shows that to a good approximation, $\nu_{c_{\text{ring}}^{\text{tb}}}$ has a dependence of the form $\sim N^2$ for large $N$.  However, the overall maximum value of $c_{\text{ring}}^{\text{tb}}$ for this system is reached when $N = 5$, reaching $\max (c_{\text{ring}}^{\text{tb}}) \approx 0.131349787116051$, which is 12\% larger than its continuous counterpart $c^{\text{cont}}_{\text{ring}}$.

\begin{figure}[H]
    \centering
    \includegraphics[width=0.45\linewidth]{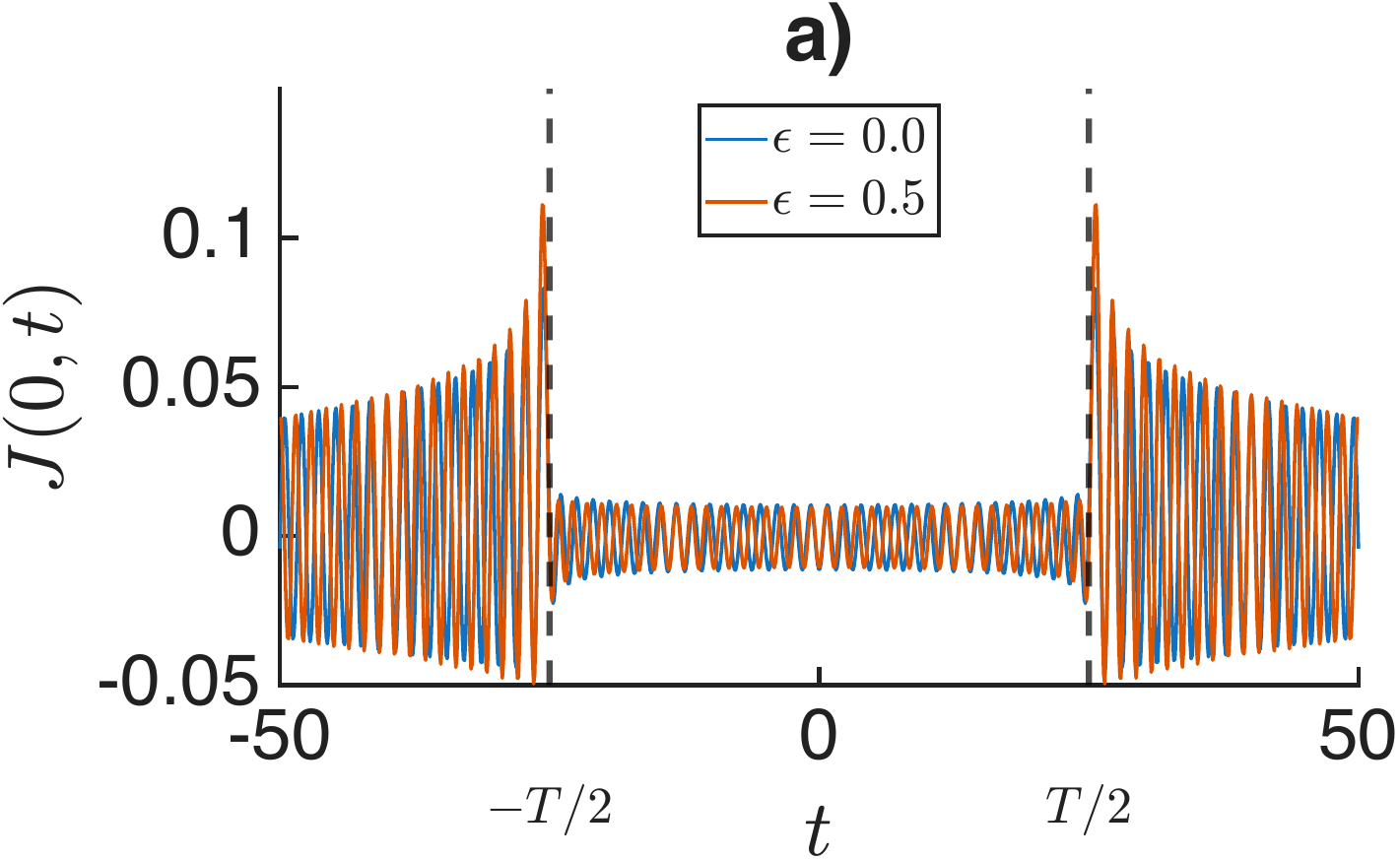}
    \includegraphics[width=0.45\linewidth]{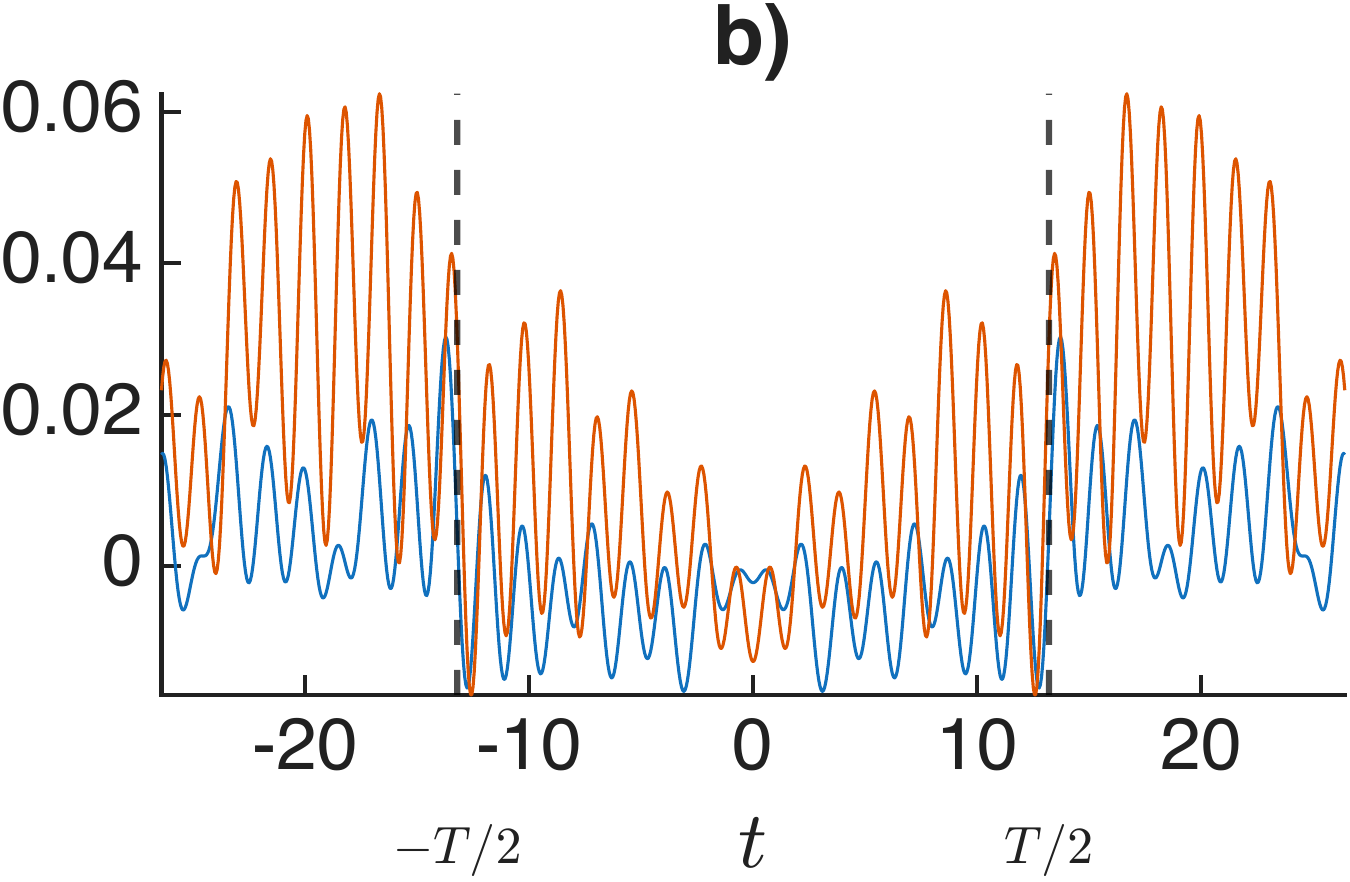}
    \caption{ Probability density flux at  $j= 0$ as a function of $t$,  for the infinite chain (a), and the periodic chain with  $N=20$ (b), with several values of $\epsilon$  using the eigenfunctions appearing in eqns.  (\ref{eqn: eig1}) and (\ref{eqn: eigen2}) respectively.  The black dashed lines indicate the limits of the integration time window.}
    \label{fig:6}
\end{figure}

In fig.~(\ref{fig:6}), we show the probability density flux at the site $j=0$ as a function of time for the states that maximize the integrated (negative) current. This was obtained by numerically evaluating the expressions in eqs.~(\ref{eqn: flux2}) and (\ref{eqn: dflux}), using the solutions of the eigenvalue equations (\ref{eqn: eig1}) and (\ref{eqn: eigen2}), respectively, for two values of $\epsilon$. In the left hand of fig.~(\ref{fig:6}), we present the infinite-chain case and set $\nu = 50$, which corresponds to the maximum value of $\nu$ used in Fig.~(\ref{fig:4}). In the right hand of fig.~(\ref{fig:6}), we display the instantaneous flux for the periodic chain with $N=20$ sites. In this case, we choose $\nu = 26.4$, which corresponds to the value at which the integrated backflow reaches $c_{\text{ring}}^{\text{tb}}$ for this lattice size.\\
 
We stress the differences between the backflow discussed in this section and that analyzed in Section \ref{sec:max}. In Section~\ref{sec:max}  we determine the largest negative instantaneous flux that can appear in these systems, whereas in the Bracken and Melloy approach, we focus on the maximum probability that flows to the left of the origin in a time interval $T$. This results from a flux oscillating with a modest amplitude around a negative value during the time window $[-T/2,T/2]$ .
\section{Conclusions and outlook}
In this work, we describe the key aspects of quantum backflow in biased tight-binding systems, for both infinite and periodic chains. We first focus on the transport properties of the system, in particular, on the significance and physical implications of the complex hopping parameter. We find that the imaginary part of the coupling plays a role analogous to a drift in diffusive systems. Furthermore, in the periodic chain case, the bias can be interpreted as a Peierls substitution, effectively incorporating the influence of a constant magnetic field into the hopping amplitudes.
Traditionally, the study of quantum backflow is based on maximizing the time integral of the probability current over a finite time window, motivated by the fact that in one-dimensional systems this quantity is bounded. This approach produces a wave function characterized by a flux oscillating closely around a negative value throughout the time window of integration, yielding a large integrated flux, though its instantaneous intensity remains relatively low. For the tight-binding systems we study, a similar behavior occurs, however, we find that the maximum value of the integrated flux is larger for the tight-binding systems than for their continuous counterparts; for example, the integrated backflow for a particle on a continuous infinite line \cite{Bracken_1994} is bounded by $c_{\text{BM}} = 0.0384517$ compared to the value $0.0764734$ that can be reached in a discrete infinite chain. In the case of periodic chains, this bound depends on the number of sites, being larger for small lattice sizes and $\epsilon = 0$. Going beyond this approach, in this paper,  we investigate the maximum values that can be reached by the instantaneous backflow amplitude at a given time and position in discrete biased tight-binding chains. We find that, for both geometries considered, the maximum backflow appears as pronounced negative flux values that decay relatively rapidly in time. This behavior is particularly evident in the infinite-chain case. By contrast, for a periodic chain,  large backflow values recur in time, although with less intensity than the maximum. 

It is worth pointing out that the  detection of backflow is still an experimental challenge, since it has not been observed in the laboratory, although some experimental proposals have been formulated \cite{Muga, Yearsley1, Barbier2021experimentfriendly}. While our approach is purely theoretical and the systems we study are extremely simple, the fact that they can sustain larger backflows than their continuous counterparts may broaden the possible experimental approaches in which this effect could be detected.  

Finally, it would be interesting to extend our analysis to backflow and its effects in other systems, including, among others, one-dimensional SSH models \cite{Wei_2025}, quantum arrival-time distributions \cite{Dhar_2015}, Bloch wave packets in periodic potentials \cite{k9km-k3dm}, disordered systems \cite{cui}, and quantum walks \cite{Zimbor}.

However, as far as we know,  in these models, even constructing the wave packets composed of non-negative momentum eigenstates necessary for studying quantum backflow, may pose a challenge.

\section*{Acknowledgments}

Francisco Ricardo Torres Arvizu acknowledges support from SECIHTI
 scholarship number 834573.  Hernán Larralde and Francisco Ricardo Torres Arvizu acknowledges the National Autonomous University of México (UNAM) through the Support Program for Research and Technological Innovation Projects (PAPIIT) number IN103724. 

\section{Appendix}
\appendix
\section{Derivation of optimal bounds on the flux for the periodic chain}
First, we define the functional from the expression for the flux at a fixed $j', t'$ and the constraint defined by the normalization condition eq.(\ref{eqn: cond1})
\begin{equation}
    \scriptstyle I_1(c_m^*, c_n)=  \frac{2 \tau \sqrt{1+\epsilon^2} }{N \hbar}\sum_{n=\eta_1}^{\eta_2} \sum_{m=\eta_1}^{\eta_2}c^*_{m}c_{n} \, e^{i\left( \frac{(2j'-1)(m-n)\pi}{N} -\frac{(E_n - E_m) t'}{\hbar} \right)} 
 \sin\left( \frac{(m + n)\pi}{N} +\xi \right) -\lambda \sum_{m=\eta_1}^{\eta_2 }  |c_{m}|^{2}.
\end{equation}
 Thus, we obtain the Euler-Lagrange equations associated with the last functional
 
     \begin{equation}
     \frac{2 \tau \sqrt{1+\epsilon^2} }{N \hbar}\sum_{n=\eta_1}^{\eta_2} c_{n} \, e^{i\left( \frac{(2j'-1)(m-n)\pi}{N} -\frac{(E_n - E_m) t'}{\hbar} \right)} 
 \sin\left( \frac{(m + n)\pi}{N} +\xi \right)=\lambda c_m,
\end{equation}
expanding we have 
\begin{equation}
\begin{split}
     \scriptstyle    \frac{2 \tau \sqrt{1+\epsilon^2}}{N \hbar}  e^{-i\left( \frac{(2j'-1)m\pi}{N} -\frac{ E_m) t'}{\hbar} \right)}\Big[ \cos\left(\frac{m}{N}\pi+\frac{\xi}{2}\right)\sum_{n=\eta_1}^{\eta_2} c_{n} \, e^{i\left( \frac{(2j'-1)n\pi}{N} -\frac{E_n t'}{\hbar} \right)} 
\sin\left(\frac{n}{N}\pi+\frac{\xi}{2}\right)\\ \scriptstyle +  \sin\left(\frac{m}{N}\pi+\frac{\xi}{2}\right) \sum_{n=\eta_1}^{\eta_2} c_{n} \, e^{i\left( \frac{(2j'-1)n\pi}{N} -\frac{E_n t'}{\hbar} \right)} 
\cos\left(\frac{n}{N}\pi+\frac{\xi}{2}\right)\Big]=\lambda c_m.
\end{split}
\end{equation}
To solve this equation, we set
 \begin{equation}
   \begin{split}
    \sum_{n=\eta_1}^{\eta_2} c_{n} \, e^{i\left( \frac{(2j'-1)n\pi}{N} -\frac{E_n t'}{\hbar} \right)} 
\sin\left(\frac{n}{N}\pi+\frac{\xi}{2}\right)=\alpha, \\ 
   \sum_{n=\eta_1}^{\eta_2} c_{n} \, e^{i\left( \frac{(2j'-1)n\pi}{N} -\frac{E_n t'}{\hbar} \right)} 
\cos\left(\frac{n}{N}\pi+\frac{\xi}{2}\right)=\beta,
\label{eqn: eee1}
\end{split}  
 \end{equation}
therefore
\begin{equation}
    c_{m}=     \frac{2 \tau \sqrt{1+\epsilon^2}}{N \hbar \lambda}    e^{-i\left( \frac{(2j'-1)m\pi}{N} -\frac{ E_m t'}{\hbar} \right)}\Big[ \alpha \cos\left(\frac{m}{N}\pi+\frac{\xi}{2}\right)   +  \beta \sin\left(\frac{m}{N}\pi+\frac{\xi}{2}\right) \Big].
    \label{eqn: sec1}
\end{equation}
Substituting eq. (\ref{eqn: sec1}) in eqns. (\ref{eqn: eee1})  we get
\begin{equation}
\begin{split}
       \frac{2 \tau \sqrt{1+\epsilon^2}}{N \hbar \lambda}   \sum_{n=\eta_1}^{\eta_2} \Big[ \alpha \cos\left(\frac{n}{N}\pi+\frac{\xi}{2}\right)  +\beta  \sin\left(\frac{n}{N}\pi+\frac{\xi}{2}\right) \Big]\sin\left(\frac{n}{N}\pi+\frac{\xi}{2}\right)=\alpha, 
\\
 \frac{2 \tau \sqrt{1+\epsilon^2}}{N \hbar \lambda}   \sum_{n=\eta_1}^{\eta_2} \Big[ \alpha \cos\left(\frac{n}{N}\pi+\frac{\xi}{2}\right)   +  \beta\sin\left(\frac{n}{N}\pi+\frac{\xi}{2}\right) \Big]\cos\left(\frac{n}{N}\pi+\frac{\xi}{2}\right)=\beta, 
\end{split}
\end{equation}
thus, we obtain the eigenvalue equation 
\begin{equation}
    \mathbf{A} \begin{pmatrix}
        \alpha\\\beta
    \end{pmatrix}= \lambda \begin{pmatrix}
        \alpha\\\beta
    \end{pmatrix}, 
\end{equation}
where the matrix $\mathbf{A}$ has the entries
\begin{align}
   A_{1,1}=A_{2,2} \scriptstyle = \frac{\tau \sqrt{1+\epsilon^2}}{N \hbar }    
\scriptstyle \csc\!\left( \frac{\pi}{N} \right) 
\sin\!\left( \frac{\pi (\eta_2+1-\eta_1)}{N} \right) 
\sin\!\left( \frac{\pi (\eta_1 + \eta_2)}{N} + \xi \right),\\
A_{1,2}=\scriptstyle\frac{\tau \sqrt{1+\epsilon^2}}{N \hbar }    [\eta_2+1-\eta_1- 
\cos\!\left( \frac{\pi (\eta_1 + \eta_2)}{N} + \xi \right) 
\csc\!\left( \frac{\pi}{N} \right) 
\sin\!\left( \frac{\pi (\eta_2+1-\eta_1)}{N} \right)],\\ 
A_{2,1}\scriptstyle =\frac{\tau \sqrt{1+\epsilon^2}}{N \hbar } [\eta_2+1-\eta_1+
\cos\!\left( \frac{\pi (\eta_1 + \eta_2)}{N} + \xi \right) 
\csc\!\left( \frac{\pi}{N} \right) 
\sin\!\left( \frac{\pi (\eta_2+1-\eta_1)}{N} \right)]
\end{align}
and its eigenvalues are
\begin{equation}
\begin{split}
 \lambda_{\pm} &=\scriptstyle\frac{\tau \sqrt{1+\epsilon^2}}{N \hbar } \Big[  \csc\!\left( \frac{\pi}{N} \right) 
\sin\!\left( \frac{\pi (\eta_2+1-\eta_1)}{N} \right) 
\sin\!\left( \frac{\pi (\eta_1 + \eta_2)}{N} + \xi \right)\\& \scriptstyle\pm \sqrt{(\eta_2+1-\eta_1)^2- 
\left(\cos\!\left( \frac{\pi (\eta_1 + \eta_2)}{N} + \xi \right) 
\csc\!\left( \frac{\pi}{N} \right) 
\sin\!\left( \frac{\pi (\eta_2+1-\eta_1)}{N} \right)\right)^2}\Big].
\end{split}
\end{equation}
The relation between $\alpha$ and $\beta$ is 
\begin{equation}
   \alpha=\pm\beta \sqrt{\frac{\eta_2+1-\eta_1- 
\cos\!\left( \frac{\pi (\eta_1 + \eta_2)}{N} + \xi \right) 
\csc\!\left( \frac{\pi}{N} \right) 
\sin\!\left( \frac{\pi (\eta_2+1-\eta_1)}{N} \right)}{\eta_2+1-\eta_1+
\cos\!\left( \frac{\pi (\eta_1 + \eta_2)}{N} + \xi \right) 
\csc\!\left( \frac{\pi}{N} \right) 
\sin\!\left( \frac{\pi (\eta_2+1-\eta_1)}{N} \right)}}.
\end{equation}
Finally, from normalization, we get
\begin{equation}
\beta = \sqrt{
\frac{
\eta_2+1-\eta_1+
\cos\!\left( \tfrac{\pi (\eta_{1} + \eta_{2})}{N} + \xi \right) 
\csc\!\left( \tfrac{\pi}{N} \right) 
\sin\!\left( \tfrac{\pi (\eta_{2}+1 - \eta_{1} )}{N} \right)
}{
 \mathcal{D_{\pm}}
}
}, 
\end{equation}
where 
\begin{equation}
\begin{split}
    \scriptstyle  \mathcal{D_{\pm}}=  (\eta_2+1-\eta_1)^{2} 
- \cos^{2}\!\left( \tfrac{\pi(\eta_{1}+\eta_{2})}{N} + \xi \right) 
\csc^{2}\!\left( \tfrac{\pi}{N} \right) 
\sin^{2}\!\left( \tfrac{\pi(\eta_2+1-\eta_1)}{N} \right) 
\\ \scriptscriptstyle\pm\csc\!\left( \tfrac{\pi}{N} \right) 
\sin\!\left( \tfrac{\pi(\eta_2+1-\eta_1)}{N} \right) 
\sqrt{(\eta_2+1-\eta_1)^{2} - 
\cos^{2}\!\left( \tfrac{\pi(\eta_{1}+\eta_{2})}{N} + \xi \right) 
\csc^{2}\!\left( \tfrac{\pi}{N} \right) 
\sin^{2}\!\left( \tfrac{\pi(\eta_2+1-\eta_1)}{N} \right)} 
\sin\!\left( \tfrac{\pi(\eta_{1}+\eta_{2})}{N} + \xi \right).
\end{split}
\end{equation}

\pagebreak
\bibliographystyle{unsrt}
\bibliography{bibliography}

\end{document}